\documentclass[aps,prd,preprint,floats,epsf,superscriptaddress,nofootinbib]{revtex4}
\usepackage[a4paper, total={7in, 9in}]{geometry} %page margin size
\usepackage{graphicx} % Required for inserting images
\usepackage{amsmath}
\usepackage{mathtools}
\usepackage{natbib,hyperref}

\begin{document}
\title{Dark $Z$-mediated dark matter with verifiable exotic scalars}

\author{Chuan-Ren Chen}
\email[e-mail: ]{crchen@ntnu.edu.tw}
\affiliation{Department of Physics, National Taiwan Normal University, Taipei, Taiwan 11677, R.O.C.}
\author{Cheng-Wei Chiang}
\email[e-mail: ]{chengwei@phys.ntu.edu.tw}
\affiliation{Department of Physics and Center for Theoretical Physics, National Taiwan University, 
Taipei, Taiwan 106319, R.O.C.}
\affiliation{Physics Division, National Center for Theoretical Sciences, Taipei, Taiwan 106319, R.O.C.}
\author{Leon M.G. de la Vega}
\email[e-mail: ]{leonmgarcia@phys.ncts.ntu.edu.tw}
\affiliation{Physics Division, National Center for Theoretical Sciences, Taipei, Taiwan 106319, R.O.C.}
\affiliation{Departamento de Física, Universidad de Sonora, 
Hermosillo, México 83000}

\date{\today}
\begin{abstract}
    In this work, we study a dark matter scenario where a dark $Z$ boson possessing mass mixing with the SM $Z$ boson couples to the DM candidate and serves as the portal to the SM. The UV origin of the mass mixing in the form of an extra dark Higgs doublet and a scalar dark singlet provides new exotic scalars which can constitute the final state of DM annihilation during freeze-out. We find that existing constraints on the observed Higgs coupling strength, exotic Higgs searches and dark matter observables complement each other, while future searches for exotic Higgs decays and resonant heavy scalars at HL-LHC will be sensitive to part of the allowed parameter space.
\end{abstract}
\maketitle
\section{Introduction}
The nature of dark matter (DM) is one of the big open questions of contemporary fundamental physics. Based on a large number of cosmological and astrophysical observations, it has been established that $\sim 80\%$ of the non-relativistic matter content of the Universe cannot be composed of Standard Model (SM) particles. While numerous experimental efforts have searched for direct signatures of DM, none have pinned down the nature of this hypothetical particle. One of the most attractive scenarios is known as the Weakly Interactive Massive Particle (WIMP), which consists of MeV-TeV scale particles thermally coupled to the SM in the early Universe. The relic density of this type of DM is determined by the annihilation rate of DM into SM particles around the time of thermal decoupling of the DM candidate \cite{Lee:1977ua}. This process, the so-called \emph{freeze-out}, ties the observed abundance of DM to the strength of the interactions between the DM and the SM fields. In turn, these DM-SM interactions can potentially lead to observable signatures in DM searches such as elastic or inelastic scattering of DM on nuclei or electrons, self-annihilation of DM in galactic halos into photons, positrons, protons or neutrinos, or production of DM in collider experiments \cite{Goodman:1984dc,Primack:1988zm}. For minimal models of WIMP DM, the combination of these constraints has severely limited the allowed parameter space of the model, relegating them to special cases where experiments are inherently not sensitive, or where there exists an enhancement of DM annihilation cross section in the early Universe while suppressing present-day observables \cite{Arcadi:2025xvj}.
A well-studied class of these cases is the $Z'$-mediated DM models, where a hypothetical neutral massive gauge boson mediates the DM-SM interactions. The $Z'$ mediator may arise from simple $U(1)$ gauge symmetries like $B$ \cite{Dulaney:2010dj,Ma:2020quj,Fornal:2020poq}, $L$ \cite{Schwaller:2013hqa,Ma:2021fre}, $B-L$ \cite{Sahu:2005fe,Khalil:2008ps,Allahverdi:2008jm,Ibe:2011hq,Alvarado:2021fbw,Chiang:2013kqa,delaVega:2021wpx}, and $L_i-L_j$ \cite{Fox:2008kb,Bell:2014tta,De:2024tvj}, among others or from gauge unification inspired models like left-right unification or $E_6$, for example. For EW-scale DM, these models are typically confined to the resonant region, where the mediator mass $m_{Z'}$ is tuned to resonantly enhance the $\text{DM}~\text{DM}\rightarrow\text{SM}~\text{SM}$ annihilation cross section in the early Universe, keeping direct, indirect and collider signatures below current limits \cite{Blanco:2019hah,Chiang:2013kqa}.
In this work, we explore a scenario where a particular $Z'$, the dark $Z$ ($Z_D$) of a dark gauge symmetry $U(1)_D$, mediates the interactions of a fermionic DM candidate with the SM, obtaining successful freeze-out while evading experimental constraints. This is done through the mass mixing of the dark $Z$ with the SM $Z$-boson, which is generated by the introduction of a second Higgs doublet, also charged under the new $U(1)_D$ symmetry. The introduction of both a second massive neutral gauge boson to the SM and a second Higgs doublet opens up new $Z_D$-mediated DM annihilation channels into the SM and exotic bosons. These channels do not need large $Z-Z_D$ mixing, which decouples DM from SM fermions, thereby avoiding a large fraction of experimental constraints on DM-SM interactions. While similar scenarios have been analyzed previously \cite{Gaitan:2014vfa,Jung:2020ukk,Jung:2023ckm,Bae:2024lov}, here we focus on the case where an additional scalar singlet is introduced. The participation of the singlet in the spontaneous dark gauge symmetry breaking opens up the mass region above 100 GeV for the new scalars and the gauge boson, which is forbidden in the scenario with only scalar doublets \cite{Jung:2020ukk,Jung:2023ckm,Bae:2024lov}. Furthermore, the singlet adds an extra degree of freedom to the dynamics of $Z-Z_D$ mixing which allows for smaller mixing angles, evading current direct collider limits on the new gauge boson parameters and reduces the DM direct detection scattering cross section. We explore the parameter space where DM freezeout is dominated by annihilations into bosons with the DM candidate mass in the 10 GeV to 5 TeV range. This scenario is previously unexplored and accommodates both DM and scalar masses in the regions where DM and LHC searches are most sensitive, respectively. The dark gauge symmetry enforces a type-I Two Higgs Doublet Model-like structure of the Yukawa couplings, leading to SM fermions coupling to exotic Higgses with a dependence on scalar mixing angles. We study possible signatures of the new scalars at high-luminosity Large Hadron Collider (HL-LHC) arising from these mixings in the resonant production of heavy scalars and in exotic decay modes of the 125 GeV Higgs boson. Ultimately, we find that the model can explain the observed relic density of dark matter and evade current constraints, while being subject to future collider probes of new scalars and gauge bosons and in DM direct detection experiments.
\section{Model}
\begin{table}[t]
    \centering
    \begin{tabular}{ccccc}
      \hline\hline
            & ~$SU(3)_C$~ & ~$SU(2)_L$~ & ~$U(1)_Y$~ & ~$U(1)_D$~\\
            \hline
      $Q$   & 3         &     2   & 1/6  & 0     \\
      $u_R$ & 3         &     1   & 2/3  & 0     \\     
      $d_R$ & 3         &     1   & -1/3 & 0     \\
      $L $  & 1         &     2   & -1/2 & 0     \\
      $e_R$ & 1         &     1   & -1   & 0     \\
      $\chi$& 1         &     1  &  0    & $Q_\chi$ \\
      \hline
      $H_1$& 1         &     2   &  1/2  & 0     \\
      $H_2$& 1         &     2   &  1/2  & 1     \\
      $\phi$& 1        &     1   & 0    & 1     \\ 
      \hline\hline
    \end{tabular}
    \caption{Field content of the model: $Q$ and $L$ are SM $SU(2)_L$ doublet quarks and leptons, respectively; $u_R$, $d_R$ and $e_R$ are SM right-handed up-type quarks, down-type quarks and charged leptons, respectively; and $\chi$ is the fermionic DM candidate.   The scalar field $H_1$ is a SM-like doublet with no dark charge, while the second doublet $H_2$ and the singlet $\phi$ have nonzero dark charge.   }
    \label{tab:fields}
\end{table}
The dark $Z$ model \cite{Davoudiasl:2012ag, Davoudiasl:2014kua} with dark matter in consideration is given by the fields in table \ref{tab:fields}. The presence of the $SU(2)_L$ doublet $H_2$ with a nonzero dark charge induces mass mixing among the SM $Z_{SM}$ boson and the dark $Z_D$ boson \cite{Lee:2013fda}.
\subsection{Scalar sector}
The most general, renormalizable, CP-conserving scalar potential allowed by gauge symmetries is given by
\begin{equation}
\begin{split}
    -V
    =& 
    \frac{\mu_1^2}{2} (H_1^\dagger H_1) + \frac{\mu_2^2}{2}(H_2^\dagger H_2) +  \frac{\mu_3^2}{2} (\phi^* \phi)  + \kappa (H_2^\dagger H_1 \phi + H_1^\dagger H_2 \phi^* )  \\
    &+ \lambda_1 (H_1^\dagger H_1)^2 + \lambda_2 (H_2^\dagger H_2)^2 + \lambda_3 (\phi^* \phi)^2 +\lambda_4 (H_1^\dagger H_1)(H_2^\dagger H_2)
    \\
    &+\lambda_5 (H_2^\dagger H_2)(\phi^* \phi)  +  \lambda_6 (H_1^\dagger H_1)(\phi^* \phi) + \lambda_7 (H_1^\dagger H_2 )(H_2^\dagger H_1)
    ~.
\end{split}
\end{equation}
We parametrize the scalar fields around their vacuum expectation values (VEVs) as follows:
\begin{equation}
    H_1=\begin{pmatrix}
       H_1^+    \\
       \frac{1}{\sqrt{2}}\left(v_1+h_1+ia_1 \right)   
    \end{pmatrix} 
    ~,~~
    H_2=\begin{pmatrix}
       H_2^+    \\
       \frac{1}{\sqrt{2}}\left(v_2+h_2+ia_2 \right)   
    \end{pmatrix} 
    ~,~~
    \phi=\frac{1}{\sqrt{2}}\left(v_\phi+h_\phi+ia_\phi \right) 
    ~.
\end{equation} 
The gauge sector restricts the scalar VEVs to satisfy
\begin{equation}
    v_1^2+  v_2^2= v_{SM}^2 = (246 ~\text{GeV})^2 
    ~,
\end{equation}
 so that the tree-level $W$ boson mass retains its SM value.
 We have defined the $H_1$ field as the only doublet coupling to SM fermions. Following this definition, we introduce the $\beta$ angle. \footnote{This definition corresponds to the standard $\tan\beta$ definition. In the usual definition of the type-I two Higgs doublet model $H_2$ couples to fermions, here $H_1$ couples to fermions, changing the labels of the Higgs doublets and resulting in this definition. }
\begin{equation}
    \tan\beta=\frac{v_1}{v_2}.
    \label{eq:tanbdef}
\end{equation}
Assuming that the three VEVs are all nonzero, the scalar spectrum after dark and EW symmetry breaking is as follows. The charged scalar sector contains two fields, one of which corresponds to a Goldstone mode $G^+$ absorbed by the $W^+$ gauge boson, while the orthogonal mode is a massive, physical field $H^+$. The neutral pseudo-scalar spectrum contains three modes, two of which are massless Goldstone bosons, $G_Z$ and $G_{Z_{D}}$, for the physical $Z$ and $Z_{D}$ gauge bosons, and a physical state $A$. These massive fields, and their masses are given by
\begin{equation}
    \begin{array}{cc}
     H^+= -\cos\beta H_1^+ + \sin\beta H_2^+
     ~,
     &
     m_{H^+}^2=\frac{1}{2} s v_{\phi}^2 -\frac{1}{2} \lambda_7 v_{SM}^2 
     ~,
     \\
     A=\frac{1}{v_A}\left(v_\phi \cos\beta a_1 -v_\phi \sin\beta a_2 +\frac{1}{2}v_{SM}\sin(2\beta) a_\phi\right)
     ~,
     &
     m_{A}^2= s \left( \frac{v_{\phi}^2}{2} + \frac{v_{SM}^2 \sin^2(2\beta)}{8} \right)
     ~,
    \end{array}
    \label{eq:scalarmasses1}
\end{equation}
where we have defined $v_A=\sqrt{v_\phi^2+v_{SM}^2\sin^2(2\beta)/4}$ and $s=\sqrt{8}\kappa/(v_\phi\sin(2\beta))$.
The neutral CP-even sector consists of three massive modes, which are defined through the diagonalization of the CP-even mass matrix $M_E$
\begin{equation}
    H_i = (\mathcal{O}^T_E )_{ij} h_j \; , \; \mathcal{O}_E^T M_E^2 \mathcal{O}_E = \text{diag}(m_{S}^2 )
    ~,
    \label{eq:scalarrotation}
\end{equation}
where $h_j=(h_1,h_2,h_\phi)$ and
\begin{equation}
    M_E^2=\begin{pmatrix}
       2 \lambda_1 v_1^2-\frac{\kappa}{\sqrt{2}}\cot\beta v_\phi & v_1v_2(\lambda_4+\lambda_7)+ \frac{\kappa}{\sqrt{2}} v_\phi & v_1v_\phi\lambda_6+ \frac{\kappa}{\sqrt{2}} v_2\\
       \cdot & 2 \lambda_2 v_2^2-\frac{\kappa}{\sqrt{2}}\tan\beta v_\phi & v_2v_\phi\lambda_5+ \frac{\kappa}{\sqrt{2}} v_1 \\
       \cdot & \cdot &  
       2 \lambda_3 v_\phi^2-\frac{\kappa}{\sqrt{2}v_\phi}v_1v_2 
    \end{pmatrix}
    ~.
    \label{eq:scalarmasses2}
\end{equation}
We parametrize the real orthogonal matrix $\mathcal{O}_E$ as a product of three Euler angle rotations 
\begin{equation}
\mathcal{O}_E=\mathcal{O}_{23}(\theta_{23})\mathcal{O}_{13}(\theta_{13})\mathcal{O}_{12}(\theta_{12}),
\end{equation}
which are defined by the scalar mixing angles $\theta_{ij}$. Throughout this work, we define $H_1$ as the SM-like Higgs boson, $m_{H_1}=125$~GeV, while $m_{H_2}$ and $m_{H_3}$ are not fixed and can be lighter or heavier than $m_{H_1}$.
\subsection{Yukawa couplings}
The Yukawa couplings of SM fermions under this setup are given by a type-I two Higgs doublet model (THDM) scenario
\begin{equation}
    \mathcal{L}_Y=Y^u\overline{Q}\tilde{H}_1u_R +Y^d\overline{Q}H_1d_R+Y^e\overline{L}H_1e_R +\text{h.c.}
\end{equation}
imposed by the $U(1)_D$ gauge symmetry instead of a discrete $Z_2$.
\subsection{Neutral gauge boson mixing}
The nonzero VEV of $H_2$, which transforms nontrivially under the weak and dark gauge groups, induces mass mixing in the neutral gauge sector. Although a bare kinetic mixing term $\epsilon F^{\mu\nu}F'_{\mu\nu}$ with arbitrary value is allowed before symmetry breaking, and $H^+$ generates a one loop contribution to the effective kinetic mixing after symmetry breaking, throughout this work we set it to zero. As we will show, the value of the mass mixing between the neutral gauge bosons depends entirely on the dynamics of gauge symmetry breaking, improving the correlation between the parameters in the gauge and scalar sectors, compared to the kinetic mixing scenario.
After performing the SM Weinberg rotation $(W_3,B)\xrightarrow{\theta_W} (A, Z_{SM})$ and identifying the massless photon, which remains unaffected in this setup, the SM $Z_{SM}$ and dark $U(1)_D$ gauge field $B_D$ possess a mass mixing term. This necessitates a further rotation, parametrized by an angle $\theta_X$, to arrive at the $Z$ and $Z_D$ mass eigenstates
\begin{equation}
    \mathcal{L}_{mix}=\frac{1}{4}\begin{pmatrix}
        Z_{SM} & B_D
    \end{pmatrix} \begin{pmatrix}
        g_Z^2 v_{SM}^2 & -2 v_2^2 g_D g_Z \\
        -2 v_2^2 g_D g_Z & 4 g_D^2 (v_2^2 +v_\phi^2)
    \end{pmatrix} \begin{pmatrix}
        Z_{SM}\\ B_D
    \end{pmatrix}, 
    \label{eq:neutralgaugemixing}
\end{equation}
where $g_Z$ and $g_D$ are the gauge coupling strengths for $Z_{SM}$ in the SM and the dark $U(1)_D$, respectively.
The mass eigenstates $Z$ and $Z_D$ are identified through the rotation 
\begin{equation}
   \begin{pmatrix}
        Z\\ Z_D
    \end{pmatrix}= \begin{pmatrix}
       \cos\theta_X & \sin\theta_X \\
        -\sin\theta_X & \cos\theta_X
    \end{pmatrix}  \begin{pmatrix}
        Z_{SM}\\ B_D
    \end{pmatrix} 
    \quad ,\quad 
    \tan2\theta_X=\frac{g_Zg_Dv_2^2}{\frac{1}{4}g_Z^2v_{SM}^2-g_D^2 (v_2^2 +v_\phi^2)}
    ~.
     \label{eq:neutralgaugerotation}
\end{equation}
The resulting fermion couplings with the neutral gauge bosons are given by
\begin{equation}
    \mathcal{L}_{V_0ff}
    = 
    - e \sum_f Q_f A_\mu \overline{f}\gamma^\mu f 
    - Z_\mu \left(
    c_X\frac{g_Z}{2}J^\mu_{NC} +s_Xg_DJ^\mu_{DC}
    \right)
    - Z_{D\mu} \left(
    -s_X\frac{g_Z}{2}J^\mu_{NC} +c_Xg_DJ^\mu_{DC}
    \right)
    ~,
\end{equation}
where $s_X(c_X)\equiv \sin\theta_X (\cos\theta_X)$, and
the neutral and dark currents are
\begin{equation}
    J^\mu_{NC}= \sum_f t^3_L(f)\overline{f}\gamma^\mu(1-\gamma_5 )f-2s_W^2 eQ_f\overline{f}\gamma^\mu f
    ~,~~ 
    J^\mu_{DC}=-Q_\chi g_D \overline{\chi}\chi
    ~.
\end{equation}
With this definition, the $Z$ boson has exactly the same fermion couplings of the SM $Z_{SM}$ boson at  tree level in the limit $\theta_X\rightarrow{0}$ (which would come from $v_2\rightarrow0$). The $Z_D$-SM fermion couplings are proportional to the $Z$ couplings, inheriting their chiral nature. This makes low-energy parity violation constraints a sensitive probe of $Z_D$ in the 0.03-10~GeV mass range \cite{Davoudiasl:2012ag, Cadeddu:2021dqx}. Lighter gauge bosons coupling to fermions have leading constraints coming from beam dump searches \cite{Filippi:2020kii,Caputo:2021eaa}, limiting $\sin\theta_X\lesssim 3 \times 10^{-8}$. For electroweak scale and heavier $Z_D$, collider searches at the LHC limit the mixing to $\sin\theta_X\lesssim 10^{-3}$ for $m_{Z_{D}}=70$~GeV, with future colliders possibly extending the mass range with this sensitivity up to $m_{Z_{D}}=500$~GeV \cite{San:2022uud}.
\subsection{Dark matter}
The DM candidate consists of the vector-like Dirac fermion field $\chi$, which only couples to the $U(1)_D$ gauge boson. Its Lagrangian can be written as 
\begin{equation}
    \mathcal{L}_\chi= \overline{\chi}\gamma^\mu
(\partial_\mu+ig_DQ_\chi B_{D\mu})\chi+m_{DM}\overline{\chi}\chi
~.
\end{equation}
In order to maintain the Dirac nature of the DM candidate after spontaneous symmetry breakdown of the gauge symmetries, we choose $Q_\chi=1/3$, which forbids the Majorana mass terms $\chi_L^C\chi_L\phi$ and $\chi_R^C\chi_R\phi$ at all orders in perturbation theory.
After the spontaneous breaking of the electroweak and dark gauge symmetries, the mass mixing of $U(1)_D$ gauge boson induces a coupling between $\chi$ and the SM-like $Z$ gauge boson. The SM fermions acquire a coupling to the $Z_D$ gauge boson as well. This mixing allows communications between $\chi$ and the fermion and scalar contents of the model. The relevant couplings for DM annihilation in the early Universe are $\overline{\chi}\chi Z(Z_D)$, $Z(Z_D) \overline{f}f$, $Z(Z_D) S V$, and $Z(Z_D) S S $, where $f$ are SM fermions, $S$ are scalars and $V$ are gauge bosons. For reference, we have collected the analytical expressions of these couplings in Appendix \ref{app:couplings}. For the freeze-out of $\chi$, we can identify four possible tree-level $2\rightarrow2$ annihilation channels:
\begin{itemize}
    \item For $m_{Z_D}<m_\chi$, $g_D \sim 1$ and $\theta_X \ll 1$, DM annihilations are dominated by the t-channel diagram into $Z_D Z_D$ pairs. This is commonly known as the \emph{Secluded WIMP} \cite{Pospelov:2007mp} limit.
    \item For $m_{Z_D(Z)} \sim 2 m_\chi$ with $\theta_X<10^{-1}$, the s-channel diagram is resonantly enhanced at the $Z_D$ or $Z$ mass, with fermion pair final states. 
    \item When $2m_{\chi}> m_V +m_{H_i}$, the scalar and gauge mixings allow Higgsstrahlung processes to dominate annihilation. If the scalar mixing and gauge mixing are negligible, only the $Z_D H_3$ final state is allowed. If the scalar mixing is large enough, $Z_D H_{1(2)}$ final states can dominate the annihilation cross section, mediated by the $Z_D$. If the gauge mixing is large enough, the $Z$ and $Z_D$ can mediate annihilations into $Z H_{1(2)}$.
    \item When $2m_{\chi}> m_A +m_{H_i}$ or $m_{\chi}> m_{H^+}$, the scalar and gauge mixings allow annihilation into two scalars. This is a $Z$/$Z_{D}$ mediated process, which can be constrained by collider searches for exotic scalar pair production, with each scalar decaying into a pair of SM fermions, or by searches for the production of a resonant pseudoscalar ($A$) decaying to the $ZH_1$ final state.
\end{itemize}
The first two scenarios have been studied previously \cite{delaVega:2022uko}, and we will thus focus on the last two scenarios. In these scenarios, the annihilation cross sections will depend strongly on the couplings of the scalars to the gauge bosons, making them susceptible to collider constraints on the scalars. 
\section{Phenomenological Constraints}
In this section, we list the phenomenological constraints to be considered for the study of the model in the parameter space defined in the next section. 
\subsection{Scalar Potential}
We impose perturbativity and potential positivity \cite{Kannike:2012pe} constraints on the scalar couplings 
\begin{equation}
    \begin{split}
        &|\lambda_i|< 4\pi\; ,\; \lambda_1>0\; ,\; \lambda_2>0\; ,\; \lambda_3>0\;,\\
        &\overline{\lambda}_{12}=\lambda_4+\theta(-\lambda_7)\lambda_7+2\sqrt{\lambda_1\lambda_2}>0\;,\\
        &\overline{\lambda}_{16}=\lambda_6+2\sqrt{\lambda_1\lambda_6}>0\;,\\
        &\overline{\lambda}_{25}=\lambda_2 +2\sqrt{\lambda_2\lambda_5}>0\;,\\
    &\sqrt{\lambda_1\lambda_2\lambda_3} + \lambda_3(\lambda_4 +\theta(-\lambda_7)\lambda_7) + \lambda_6( \sqrt{\lambda_1} +\sqrt{\lambda_2} ) + \sqrt{\overline{\lambda}_{12}\overline{\lambda}_{16}\overline{\lambda}_{25} }>0\;,
    \end{split}
\end{equation}
where the Heaviside step function $\theta(x)=1$ for $x\geq 0$ and $\theta(x)=0$ otherwise. \\
Additionally, we impose tree-level perturbative unitarity constraints on the scalar couplings \cite{Muhlleitner:2016mzt}
\begin{equation}
    \begin{split}
        &\left|\frac{1}{2}\left(\lambda_1+\lambda_2\pm\sqrt{(\lambda_1-\lambda_2)^2+4\lambda_7^2}\right)\right|< 8\pi \; ,\\
        &\left|\frac{x_i}{4}\right| < 8\pi \; , \; i=1,4
        ~,
    \end{split}
\end{equation}
where $x_i$ are the three roots of the polynomial 
\begin{equation}
    a_0+a_1 x +a_2 x^2+ a_3 x^3 
\end{equation}
with
\begin{equation}
    \begin{split}
        & a_0= 64 (6\lambda_5^2\lambda_1+ 6\lambda_6^2\lambda_2-9\lambda_1\lambda_2\lambda_3-8\lambda_4\lambda_5\lambda_6+4\lambda_4^2\lambda_3-4\lambda_5\lambda_6\lambda_7+\lambda_7^2\lambda_3)\;,\\
        & a_1= 16(-2\lambda_6^2-2\lambda_5^2+3\lambda_1\lambda_3+3\lambda_2\lambda_3+9\lambda_1\lambda_2-4\lambda_4^2-4\lambda_4\lambda_7-\lambda_7^2 )\;,\\
        & a_2=-4\lambda_3-12\lambda_1-12\lambda_2\;,\\
        &a_3=1.
    \end{split}
\end{equation}
\subsection{125-GeV scalar and exotic Higgs physics}
The scalar states in the theory couple to the SM gauge bosons and fermions. One of the neutral scalar bosons must possess all the properties of the 125-GeV scalar boson observed by CMS \cite{CMS:2022dwd} and ATLAS \cite{ATLAS:2022vkf}. These include the measured Higgs coupling strengths to fermions ($\kappa_f$), to $W$ and $Z$ bosons ($\kappa_{V}$), to gluons ($\kappa_g$) and to the photon ($\kappa_\gamma$), as well as the limits to the exotic decay modes of the Higgs boson. We have collected the analytical expressions of the $H_1$ tree-level couplings in Appendix \ref{app:couplings}. Since no other fundamental scalar has been observed in collider experiments, the rest of the neutral scalars and the charged scalars must evade the limits on their production cross sections and decay branching ratios set by LEP, ATLAS and CMS. An additional important indirect constraint on the scalar spectrum comes from the charged Higgs contribution to flavor-violating decays, in particular to $B\rightarrow X_s\gamma$, from which the type-I THDM disfavors $\tan\beta<1.5-2$, depending on the value of $m_{H^+}$ \cite{Arbey:2017gmh}.

\subsection{$Z_D$ quark lepton interactions}

The $Z-Z_D$ mixing induces small $Z_D f f$ interactions, where $f$ is a SM quark or lepton. This makes the $m_{Z_D}-\sin\theta_X$ parameter space susceptible to limits from new gauge boson searches in atomic parity violation (APV), beam dump, $e^+e^-$ and hadron colliders.  To implement these constraints, we use the \texttt{DarkCast} code \cite{Ilten:2018crw,Baruch:2022esd} to recast existing constraints on dark photons searches to the dark $Z$ model. The leading constraints in the $Z_{D}$ mass regime of interest in this work, $\sim 10$~GeV, are the searches for on-shell production of the $Z_D$ with visible decay products at LHCb \cite{LHCb:2019vmc}. 
\subsection{Oblique parameters}
The gauge and scalar sectors of the model induce an effect on the value of the oblique parameters $S$ and $T$ that measure deviations from the SM vacuum polarization functions of electroweak gauge bosons. In this model, the leading correction from the dark $Z$ gauge boson mass mixing with the SM $Z$ gauge boson arises at the tree level, while the scalar sector induces one-loop corrections. We consider only the pure gauge \cite{Babu:1997st,Davoudiasl:2023cnc,Cheng:2024hvq} and pure scalar \cite{Grimus:2008nb} contributions, neglecting any potential one-loop gauge-scalar contribution, as they are suppressed by both the loop factor and the small gauge mixing angle. The most stringent experimental limits on these parameters are set by global fits of electroweak precision data \cite{ParticleDataGroup:2024cfk}
\begin{equation}
    S=-0.05\pm 0.07\; ,\; T=0.00\pm 0.06 \; ,\; \rho_{ST}=93\%,
\end{equation}
where $\rho_{ST}$ is the correlation between these observables. We have taken $U=0$ as it is suppressed in comparison to $S$ and $T$. We consider a benchmark point acceptable if the predicted values of $S$ and $T$ are not excluded at the $90\%$ CL, given by
\begin{equation}
    \chi^2_{ST}
    =
    \begin{pmatrix}
        S_{th}+0.05\\T_{th}
    \end{pmatrix}^T \left[ \begin{pmatrix}
        0.07 &0\\
        0 &   0.06
    \end{pmatrix}\begin{pmatrix}
        1 &0.93\\
        0.93 &   1
    \end{pmatrix} \begin{pmatrix}
        0.07 &0\\
        0 &   0.06
    \end{pmatrix}\right]^{-1}
    \begin{pmatrix}
        S_{th}+0.05\\T_{th}
    \end{pmatrix} < 4.6
    ~.
\end{equation}
\subsection{SM boson decays}
The introduction of new light bosons to the SM opens up new decay channels for heavy SM particles. In this model, the novel decay channels which can have large contributions to the $Z$ and $H_1$ decay widths are
\begin{equation}
\begin{split}
    Z\rightarrow& Z_{D}A, Z_{D}H_2, Z_{D}H_3, H_2A, H_3A 
    ~,
\\
    H_1\rightarrow& AA, Z_{D}Z_{D}, Z Z_{D}, ZA, Z_{D}A, H_2H_2, H_3H_3, H_2H_3
    ~.
 \end{split}
\end{equation}
The width of the $Z$ boson has been precisely measured at LEP \cite{Janot:2019oyi}
\begin{equation}
\Gamma(Z)=2.4955\pm 0.0023 \text{ GeV}.
\end{equation}
In order not to saturate the precise limits on the $Z$ boson properties at $3\sigma$, we impose a limit on the total BSM decay width $\Gamma_{BSM}(Z)$
\begin{equation}
   \Gamma_{th}(Z)=\Gamma_{SM}(Z) +\Gamma_{BSM}(Z) <2.4955+3\times( 0.0023) \text{ GeV}. \\
\end{equation}
On the other hand, the width of the SM-like Higgs boson observed at the LHC has only been obtained indirectly, which yields an experimental value consistent with the SM expectation but with a large uncertainty \cite{ParticleDataGroup:2024cfk}
\begin{equation}
    \Gamma(H_1)=3.7^{+1.9}_{-1.4} \text{ MeV}.
\end{equation}
We impose similarly  
\begin{equation}
   \Gamma_{th}(H_1)=\Gamma_{SM}(H_1)+ \Gamma_{BSM}(H_1) <3.7+3\times( 1.9) \text{ MeV}. \\
\end{equation}
The daughter particles of BSM $H_1$ decays commonly decay promptly to $bb$ or $\tau\tau$ pairs, which can further constrain the allowed parameter space with results from $H_1\rightarrow 4b, 2b+2\tau$ searches. We impose these limits using the \texttt{HiggsBounds} code.
\subsection{Dark Matter}
\begin{figure}
    \centering
    \includegraphics[width=0.6\linewidth]{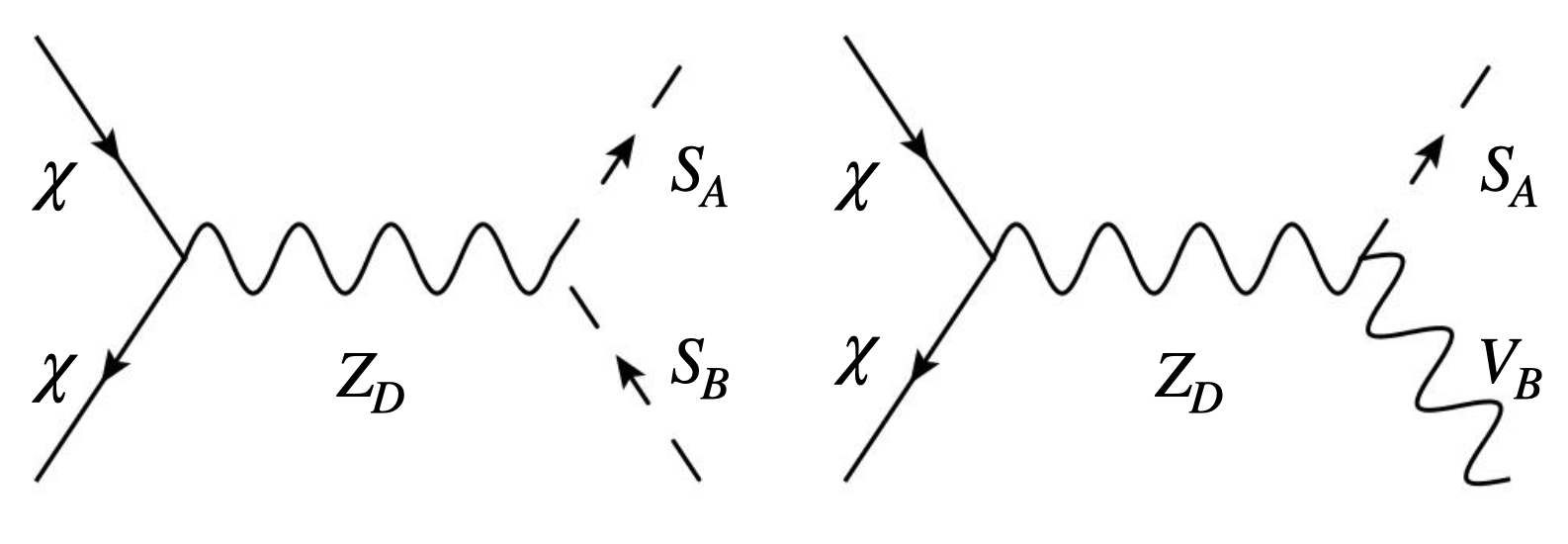}
    \caption{Dark matter annihilation channels. The left channel is the leading diagram for annihilation into two scalars. The possible combinations of $S_A,S_B$ are $\{H^+H^-,AH_i\}$.  The diagram on the right side is the leading contribution to annihilation into a scalar and a vector boson. The possible combinations of $S_A,V_B$ are $\{H^+W^-,H_iZ,H_iZ_D\}$.}
    \label{fig:dmannihilation}
\end{figure}

The interactions of DM $\chi$ with $Z_D$ provide an avenue for interactions between the dark sector and SM sector. Considering $\chi$ as a thermally produced particle, we calculate the freeze-out relic density of $\chi$ through the processes depicted in figure~\ref{fig:dmannihilation}.
These processes should drive the freeze-out of the DM candidate to its observed relic density \cite{Planck:2018vyg}
\begin{equation}
    \Omega_{DM}h^2=0.1198\pm0.0012\;.
\end{equation}
The $Z-Z_D$ mixing induces tree-level elastic scattering between DM and a nucleon, which is probed in direct detection experiments. In this setup, we potentially obtain both spin-independent and spin-dependent cross sections with both proton and neutron. These cross sections have been constrained by numerous experiments. We take these limits from currently leading collaborations, PICO-60 \cite{Amole:2019}, XENONnT \cite{XENON:2023cxc} and LZ \cite{LZ:2024zvo}.
\section{Numerical Analysis}
In the model under consideration, the presence of the complex scalar singlet charged under the dark $U(1)_D$ symmetry allows for the massive physical pseudoscalar field to be similar, in fermion and SM gauge couplings, to the type-I THDM pseudoscalar. This can be accomplished in the $v_\phi>v_{SM}$ regime, where most of the $Z_D$ Goldstone is constituted by $a_\phi$. In this regime, we observe large $Z_D-A-H_i$ couplings, with $H_i$ being a CP-even scalar. This opens up the possibility of obtaining a successful freeze-out from $Z_D$-mediated scalar channels without the need for a large neutral gauge mixing angle. In turn, this can explain the absence of a direct detection signal of DM, as a small $\theta_X$ angle suppresses this cross section. We have performed a scan of the model parameters:
\begin{equation}
    \{m_A,m_{H^+},m_{H_2},m_{H_3},\theta_{ij},v_\phi,g_D,\tan\beta,m_{\chi}\}
    ~,
\end{equation}
from which we obtain analytically and numerically the scalar potential couplings using Eqs.~\eqref{eq:scalarmasses1}-\eqref{eq:scalarmasses2} and \eqref{eq:neutralgaugemixing}, \eqref{eq:neutralgaugerotation}. Concentrating on the DM annihilation into bosons, we impose the conditions
\begin{equation}
    m_A+m_S<2m_{\chi}<2m_{Z_D}\; \text{     or    } \;2m_{H^+}<2m_{\chi}<2m_{Z_D}\, \text{     or    } \; m_V+m_S<2m_{\chi}<2m_{Z_D}\;, 
\end{equation}
where $S=H_i,H^+$ and $V=Z,Z_D,W^+$, considering only electrically neutral combinations of particles.
The resulting benchmark points are subject to the analytical perturbativity, perturbative unitarity, scalar potential positivity and electroweak oblique parameter constraints. We have implemented the model in \texttt{SARAH}  \cite{Staub:2013tta}, \texttt{micrOMEGAs} \cite{Alguero:2023zol}, \texttt{HiggsTools} \cite{Bahl:2022igd} and \texttt{DarkCast} \cite{Ilten:2018crw,Baruch:2022esd} in order to set bounds on DM observables, scalar couplings, cross sections and dark $Z$ searches.
\section{Results}
\begin{figure}
    \centering
    \includegraphics[width=0.5\linewidth]{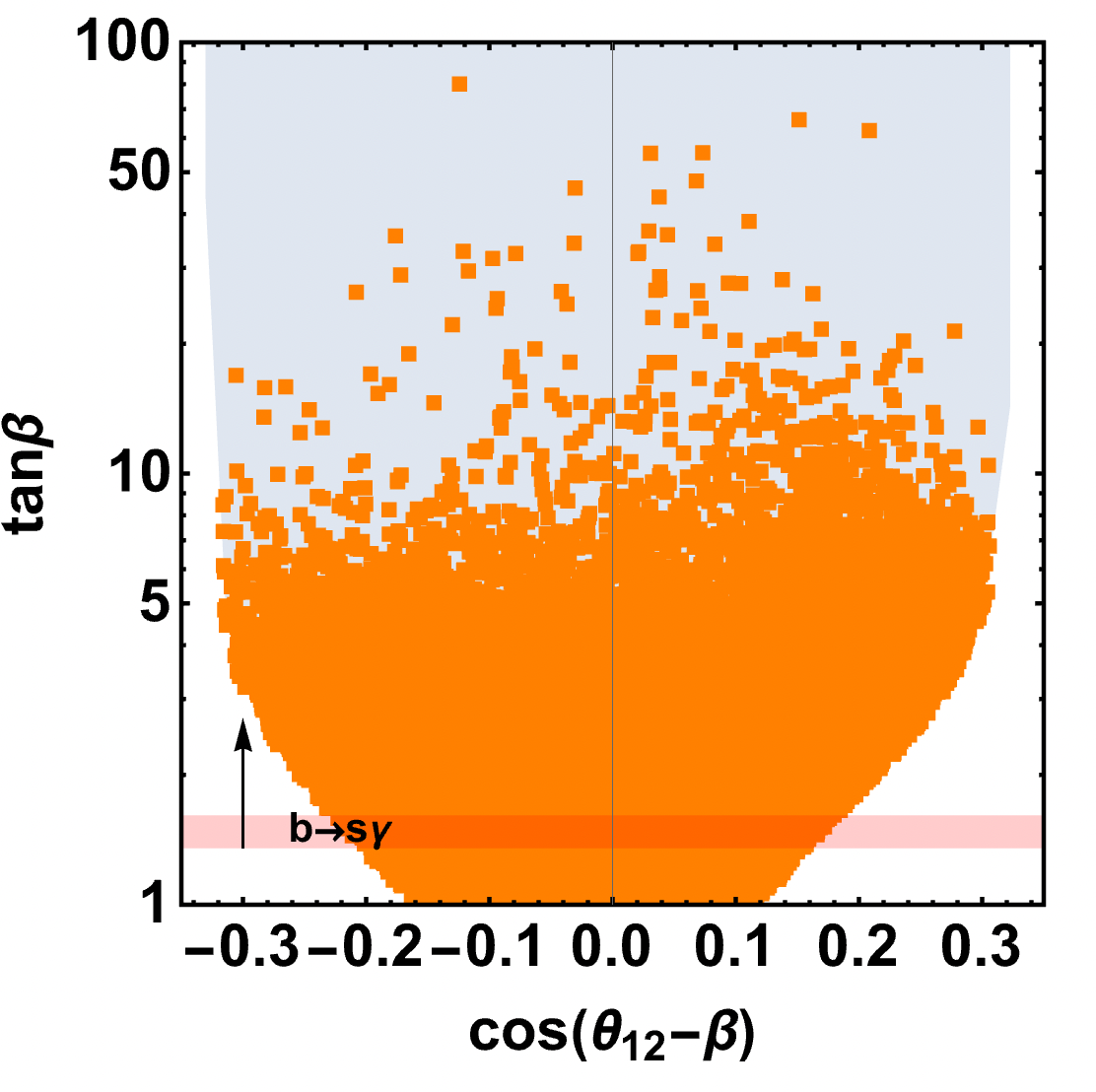}
    \caption{Allowed parameter space in the plane of $\tan\beta$ and $\cos(\theta_{12}-\beta)$. The gray region corresponds to points satisfying the $Z$ boson mass, perturbative scalar couplings, scalar potential positivity and SM Higgs-like couplings constraints. Orange points additionally satisfy tree-level perturbative unitarity constraints of the scalar couplings. The red horizontal band represents the limits from charged Higgs-induced flavor-violating processes. The region below the band is excluded for any charged Higgs mass, while points inside the band can be allowed if the charged Higgs is heavy enough to avoid the constraint. Points above the band are allowed for the masses obtained in our scan. }
    \label{fig:costhvstan}
\end{figure}
First, we perform a scan over the considered parameter space. As we are interested in studying the previously unexplored scenario where DM freezeout proceeds through annihilations into scalar bosons which can be searched for at the LHC, we focus on the region of the parameter space with new scalars in the 10 GeV to 1 TeV region. We set the following limits on the free parameters:
\begin{equation}
\begin{split}
    m_A\; \in \; (10,1000) \text{ GeV}\;,\;&
    m_{H^+}\; \in \; (100,1000) \text{ GeV}\;,\;   m_{H_2}-m_{H^+}\; \in \; (-10,10) \text{ GeV}\\
   m_{H_3}\; \in \; (100,1000) \text{ GeV}\;,\;& \tan\beta\; \in \; (0.1,100)\;,\; v_\phi\; \in \; (10,1000) \text{ GeV}\\
    & \theta_{ij}\; \in \; (-\pi/2,\pi/2)\; , 
    \end{split}
    \label{eq:scanlimits}
\end{equation}
where we have imposed an approximate twisted custodial symmetry $m_{H^+}\sim m_{H_2}$ in order to avoid saturating the constraints from the electroweak oblique parameters \cite{Gerard:2007kn}. Notice that in this model, the custodial symmetry is ``approximate'' even in the limit of mass equalities as $H_2$ has a nonzero singlet component. This is also true of the other custodial symmetry $m_A\sim m_{H^+}$ in this case because $A$ possesses a singlet component. With this consideration, oblique parameter constraints will not be automatically satisfied as there will be contributions from scalar mixing and $Z-Z_{D}$ mixing. We have chosen the twisted custodial symmetry in order to allow the pseudoscalar $A$ to be light. We are interested in the possibility of a light pseudoscalar $A$ so that the decay channels $H_1\rightarrow H_3H_3,AA$ are kinematically open. This leads to the SM-like Higgs decay channel into scalars in the $2b2\tau$ detection channel (which is dominated by scalar mixing from the scalar couplings) and the same for the pseudoscalars (which is dominated by pseudoscalar mixing, and hence depends only on the values of the VEVs).
From the obtained benchmark points, we select those that reproduce correctly the $Z$ mass, lead to perturbative scalar couplings, and a positive scalar potential. Furthermore, we require that the properties of the observed Higgs $H_1$ remain close to their experimental limits. Using \texttt{HiggsSignals}, we select the points which are allowed at $95\%$ CL. We plot the parameter space allowed by these constraints as the gray region in figure~\ref{fig:costhvstan}, alongside the flavour constraints from charged Higgs loops as the red horizontal band. The perturbative unitarity constraints significantly reduce this space, as shown by the remaining orange points in the same figure, showing that the region with $v_2>v_1$ is disfavored. 
\begin{figure}
    \centering
    \includegraphics[width=0.55\linewidth]{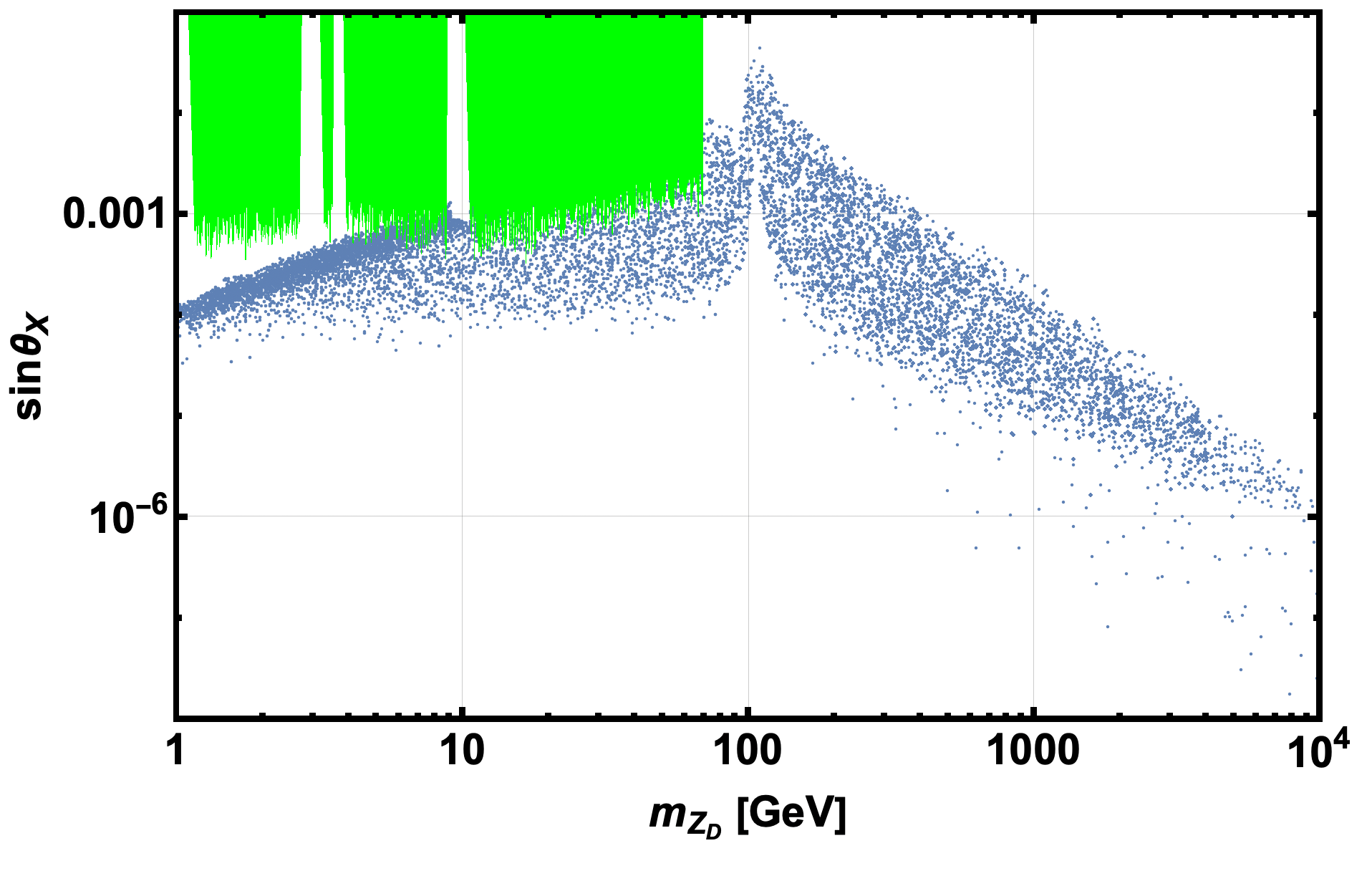}
    \includegraphics[width=0.4\linewidth]{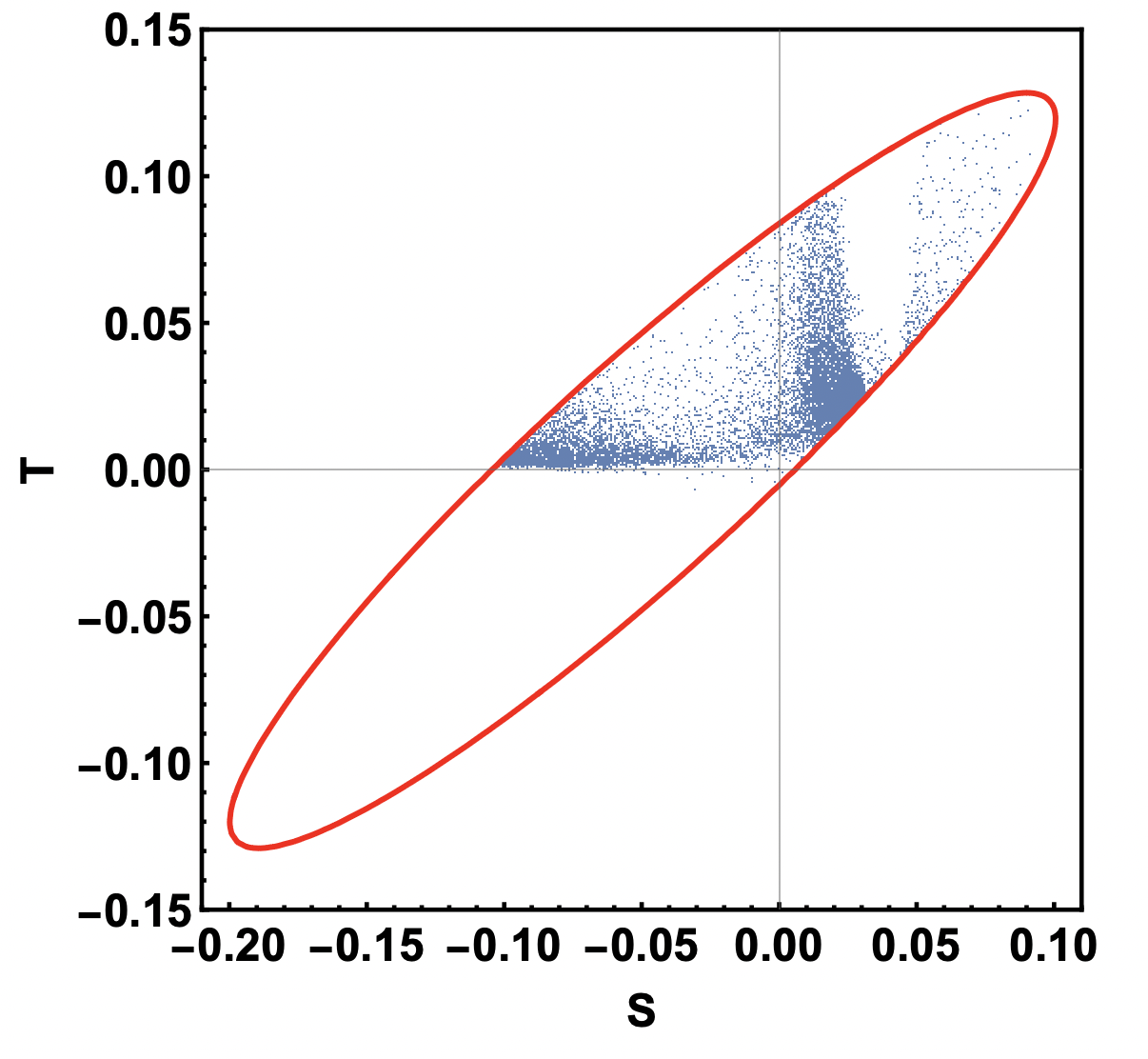}
    \caption{Left: Parameter space in the $m_{Z_D}-\sin\theta_X$ plane of the scan defined in eq.~\ref{eq:scanlimits} in blue, together with the limits from dark photon searches at LHCb \cite{LHCb:2019vmc} recast to the dark $Z$ model \cite{Ilten:2018crw,Baruch:2022esd} in green. Right: Oblique parameter constraints on the generated benchmark points. The red oval corresponds to the $90\%$ CL contour of the global fit \cite{ParticleDataGroup:2024cfk} when taking $U=0$.}
    \label{fig:MZDvsthetaX}
\end{figure}
Next, we impose the electroweak oblique parameter constraints and collider limits on $Z_{D}$ production at the LHC. We show the remaining parameter space in the $m_{Z_{D}}-\theta_X$ plane in figure \ref{fig:MZDvsthetaX}, with the LHC constraints from $Z_D$ searches overlaying in green. We notice a striking difference between this model and the minimal Dark $Z$ model (see figure 1 of \cite{Jung:2023ckm}) in the relaxed freedom in these two parameters obtained by the introduction of the singlet. In the minimal model with only doublets, there is an almost one-to-one correspondence between $m_{Z_{D}}$ and $\theta_X$, as the $Z-Z_{D}$ mixing and masses depend on a single parameter, $\tan\beta$. Additionally, in the model considered here, a $Z_{D}$ heavier than the electroweak scale is achievable, although with diminishing mixing. In the right-hand side of figure~\ref{fig:MZDvsthetaX}, we show the remaining points projected on the $S-T$ oblique parameter 2-D plane. We show in red the experimental constraint on these parameters which exclude points at the 90$\%$ CL. We see that even with the imposition of twisted custodial symmetry, the contributions to these parameters from gauge and scalar boson mixings can drive them significantly away from zero.\\
In figure \ref{fig:scalarspectrum} we show the scalar spectrum after the imposition of the previous constraints. We note that we can achieve charged Higgs masses in the $m_{H^+}\in   (100,700)$ GeV range, pseudoscalar masses in the $m_A\in (10,700)$ GeV range. For masses above $300$ GeV, we see that $m_A$ and $m_{H^+}$ tend to be well correlated. This is expected, as masses larger than the EW scale for either one requires a larger $\kappa$ parameter of the scalar potential, which in turn raises the other mass  (see eq. \ref{eq:scalarmasses1}). For the $H_3$ mass, we observe that masses in the whole scanned range $m_{H_3}\in(10,1000)$ GeV are achievable with little correlation to the charged Higgs mass.
\begin{figure}
    \centering
    \includegraphics[width=0.45\linewidth]{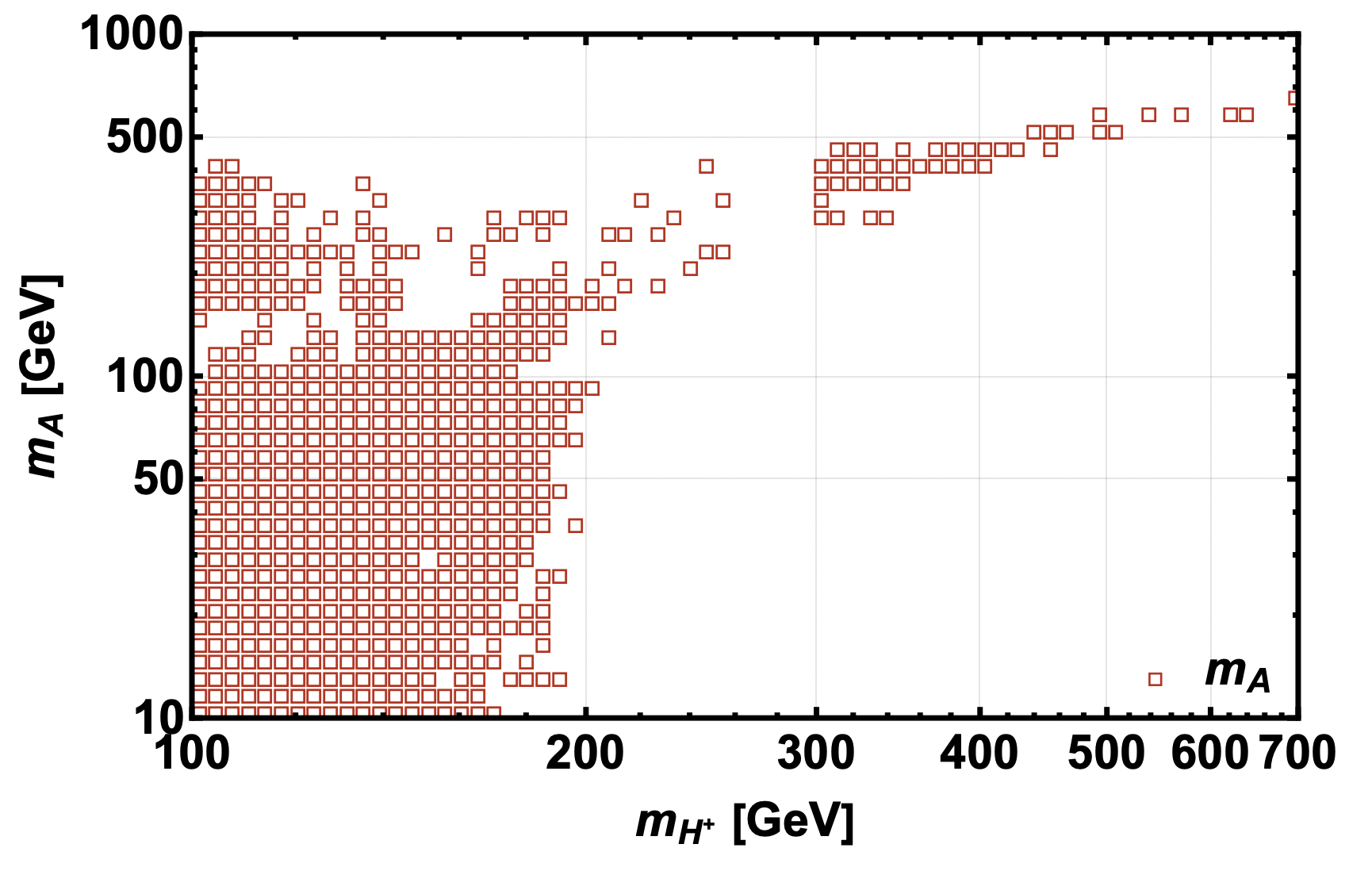}
    \includegraphics[width=0.45\linewidth]{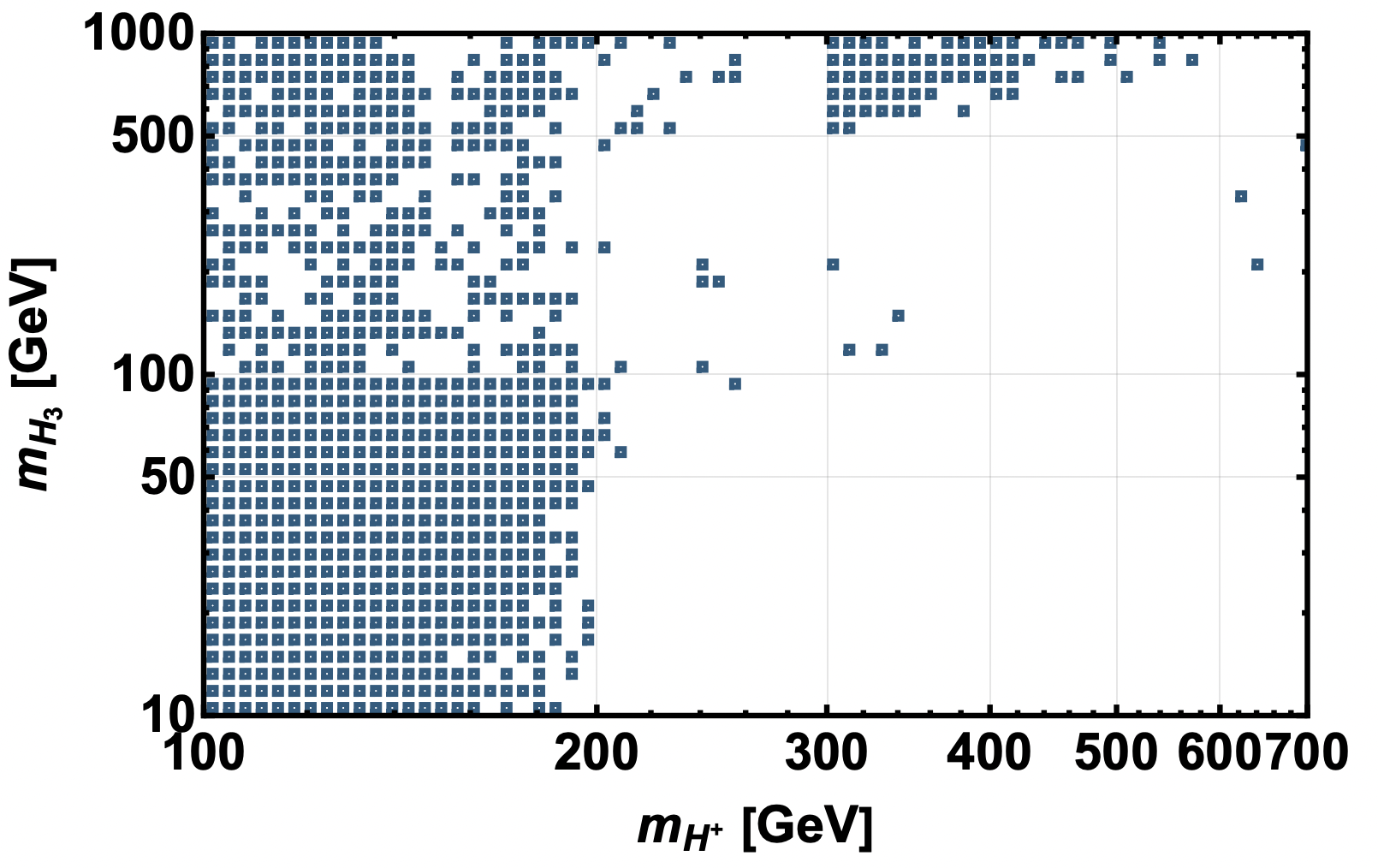}
    \caption{Scalar spectrum of generated points. These points satisfy perturbativity, perturbative unitarity, scalar positivity, $H_1$ SM-like properties, oblique parameters and $Z$ and $H_1$ decay width constraints. Left: Scalar spectrum in the $m_{H^+}-m_A$ plane. Right: Scalar spectrum in the $m_{H^+}-m_{H_3}$ plane. For all points, $m_{H_2}\approx m_{H^+}$.}
\label{fig:scalarspectrum} 
\end{figure}
Now, we impose the limits from searches for exotic Higgses at colliders and from the precise $Z$ decay width determination using \texttt{HiggsBounds} and CalcHEP, respectively. With the remaining points in the parameter space, we perform a scan with \texttt{micrOMEGAs}, imposing the kinematic constraints forbidding secluded annihilation of the DM candidate into the dark $Z$ and allowing at least one of the diboson final states of $SS$ or $SV$ with $S=H_{i},A,H^+$ and $V=Z,Z_{D},W^+$. We show the points with viable DM relic density $\Omega_\chi h^2=0.1198\pm5\%$ in figure~\ref{fig:dmpheno}. On the left-hand side, we show the $Z_{D}$ mass plotted against the DM mass, showing remarkably that points away from the resonance $2m_\chi\sim m_{Z_{D}}$ are viable, in stark contrast to the minimal dark $Z$ model. This is again due to the tight relationship between the $Z_{D}$ mass and the gauge mixing angle in the minimal model, which makes the heavier $Z_{D}$ acquire larger mixing. This in turn drives the t-channel $Z_{D}$-mediated direct detection cross section larger than the experimental limits, unless the s-channel annihilation cross section is enhanced by the $Z_{D}$ resonance. On the right-hand side, we plot the spin-independent elastic scattering cross section against the DM mass, along with current~\cite{LZ:2024zvo} and future limits~\cite{XLZD:2024nsu}, as well as the neutrino floor for Xenon~\cite{OHare:2021utq}. We see that although direct detection constraints are important for a sub-$100$-GeV DM candidate, there are still viable points in this region and that, as expected, the larger the DM mass is the lower the direct detection cross section becomes. This behavior is understandable by the forbidding of the secluded regime $m_\chi<m_{Z_{D}}$ together with the $m_{Z_{D}}-\theta_X$ dependence of the model (see figure~\ref{fig:MZDvsthetaX}). Briefly, the kinematic conditions require $m_{Z_{D}}$ to grow as $m_\chi$ grows. Above $\sim100$~GeV, $\theta_X$ tends to decrease as $m_{Z_{D}}$ grows, which suppresses the direct detection cross section.
 It is worth noting that points below the projected XLZD sensitivity are already below the Xenon neutrino floor, inside the so-called neutrino fog. Their detectability will depend on technology improvements and the reduction of the neutrino background uncertainties, among other factors~\cite{OHare:2021utq}.
\begin{figure}
    \centering
    \includegraphics[width=0.48\linewidth]{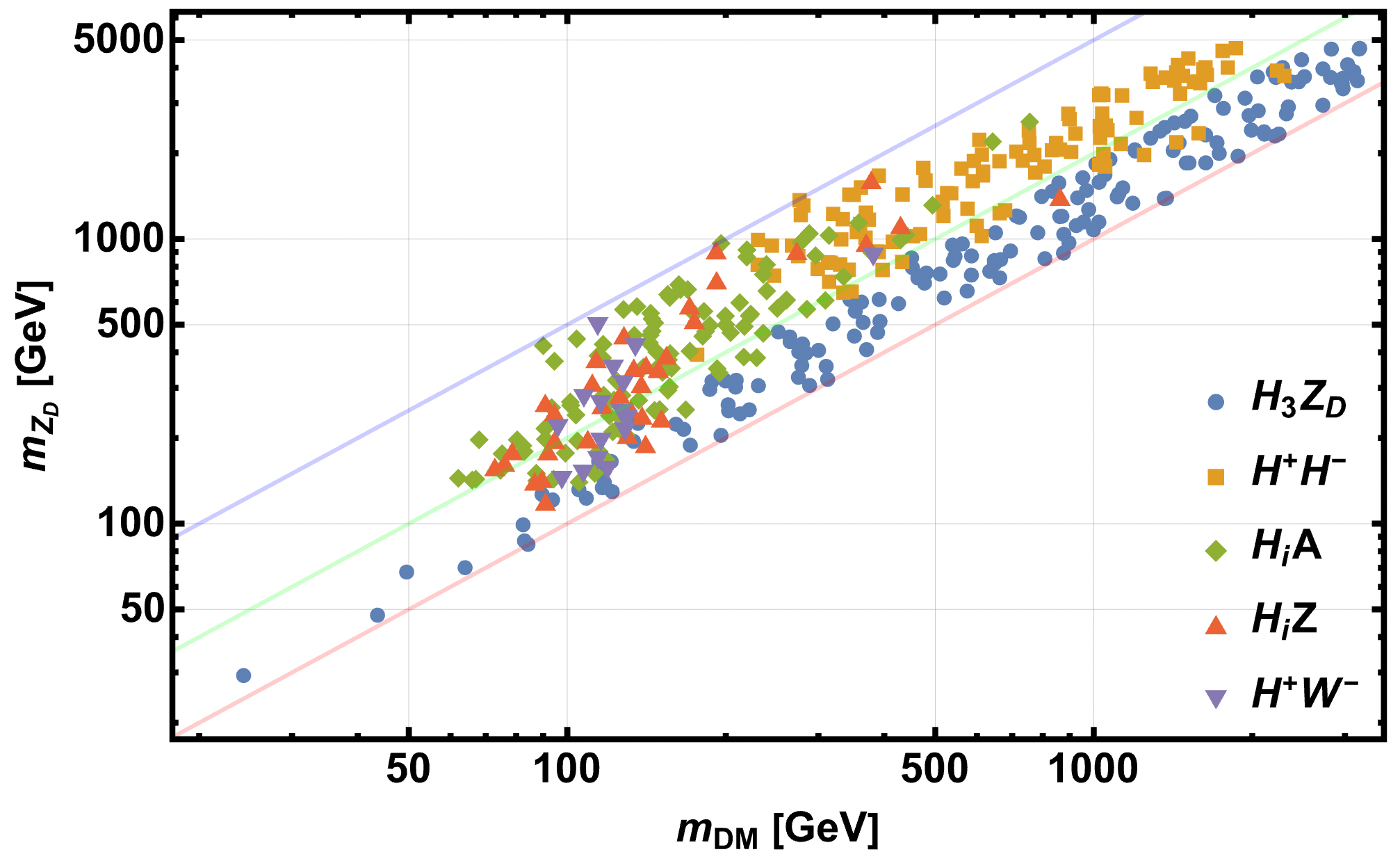}
    \includegraphics[width=0.48\linewidth]{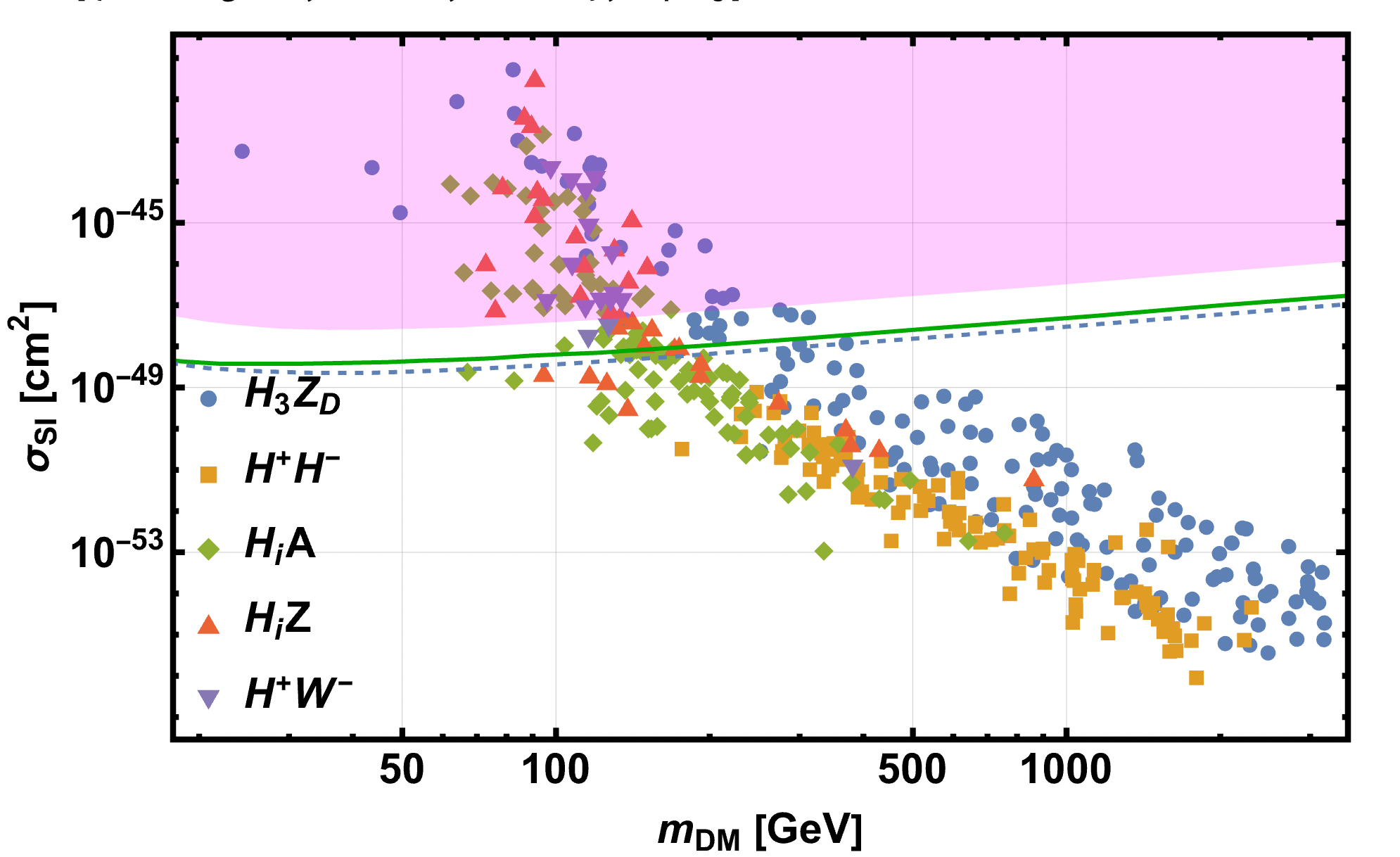}
    \caption{Dark Matter constraints on $m_{DM}$ and $m_{Z_D}$. Left: Benchmark points satisfying theoretical, electroweak, Higgs signal strengths, exotic scalar and relic density constraints in the $m_{DM}-m_{Z_D}$ plane. Blue, green and red diagonal lines correspond to $m_{Z_D}/m_{DM}=5,\,2$ and $1$, respectively. Right: Same set of benchmark points in the $m_{DM}-\sigma_{SI}$ plane. Magenta region is excluded by the latest results of the LZ collaboration~\cite{LZ:2024zvo}. Blue dashed line corresponds to projected XLZD limits for 200 ton-years~\cite{XLZD:2024nsu} of exposure, representative of a conservative projection for next-generation experiments. In solid green, we show the neutrino floor for Xenon~\cite{OHare:2021utq}, the material used in both LZ and XLZD. In both panels, we show the final state of the largest annihilation cross section at freeze-out according to the shape of the point as indicated in the insets. Here $H_i=H_1,\,H_2$ or $H_3$.}
    \label{fig:dmpheno}
\end{figure}
Finally, we turn to the prospect of detecting signals of the new scalars introduced in this model. The exotic scalars are crucial for DM phenomenology, and we have taken care not to decouple them completely from the SM. Detecting an exotic scalar in a collider search would constitute an important complementary probe of the model and in some cases the leading possibility, as direct detection probes of DM are disfavored for heavy dark matter.
We consider two possible scenarios for exotic scalar detection. The first consists of light exotics $S$ that are produced from the SM-like Higgs decay. The two possible candidates, as discussed before, are $S=H_3$ and $S=A$. One of the most promising detection channels of the process at the LHC for $H_1\rightarrow SS$ is the $2b2\tau$ final state due to its relatively small background, compared to $4b$. For points that satisfy all previous constraints and $2m_S<m_{H_1}$, we use CalcHEP to calculate the branching ratios $BR(H_1\rightarrow SS)$ and $BR(S\rightarrow \overline{b}b(\overline{\tau}\tau))$ and compare them to the sensitivity projections of HL-LHC to this process \cite{Cepeda:2019klc}. In figure~\ref{fig:lighthiggsHLLHC}, we show the points in the $m_S-BR(H_1\rightarrow SS\rightarrow \overline{b}b\overline{\tau}\tau)$ plane, along with the sensitivity projection for the full run of the HL-LHC (3000~$\text{fb}^{-1}$). We see that a significant amount of the points lie within the sensitive region, while some have suppressed branching ratios and are unobservable. 
\begin{figure}
    \centering
    \includegraphics[width=0.5\linewidth]{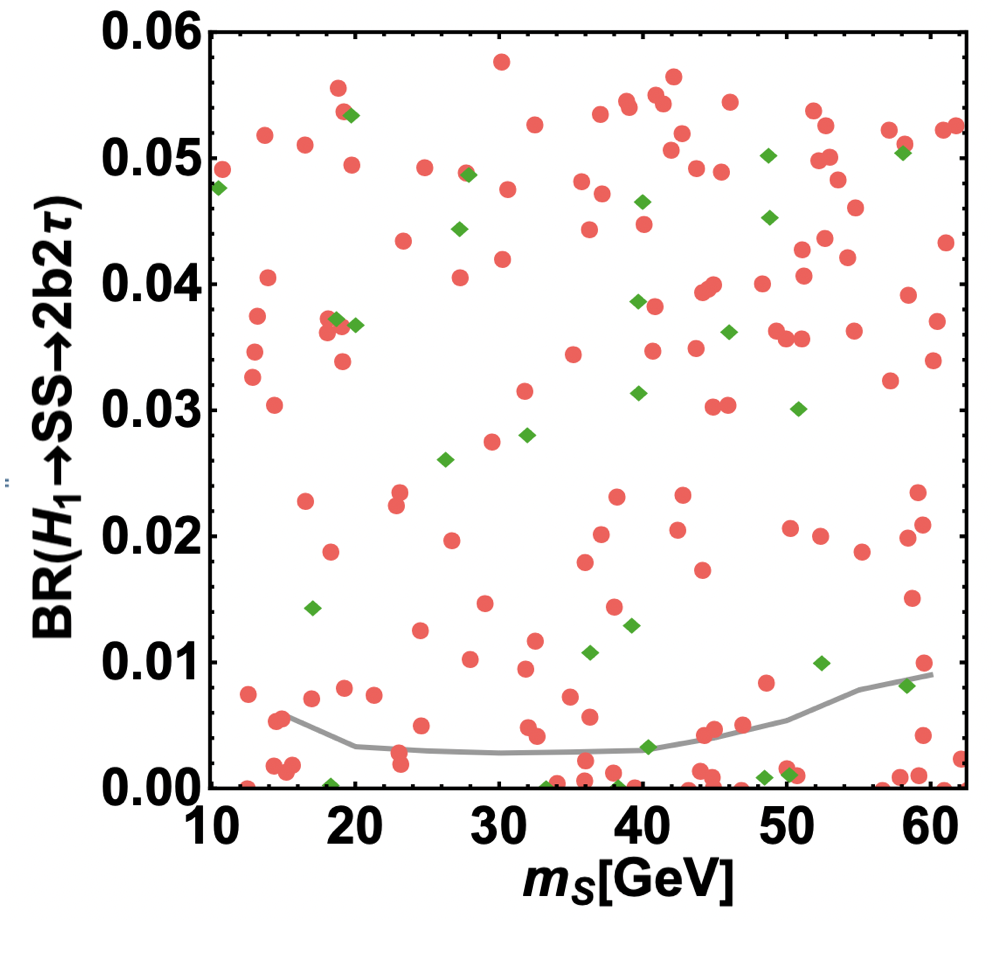}
    \caption{Projections for the observation of the exotic decay $H_1\rightarrow2b2\tau$ at the HL-LHC, mediated by the intermediate processes $H_1\rightarrow S S\rightarrow2b2\tau$, with $S=H_3,A$. Red circles indicate the $H_1\rightarrow H_3 H_3$ benchmark points, while green diamonds indicate $H_1\rightarrow A A$. We show as a gray line the sensitivity of HL-LHC at full luminosity \cite{Cepeda:2019klc}.}
    \label{fig:lighthiggsHLLHC}
\end{figure}
The presence of an extra scalar singlet in the THDMs can significantly enhance the branching ratios of heavy scalars into $H_1$ pairs or into a $Z$ and a $H_1$ \cite{Baum:2018zhf}. For this reason, the second scenario we consider for exotic scalar detection at the HL-LHC is with heavy resonant scalars decaying in these channels. We consider a gluon-fusion production of a scalar $H_2$, which then decays into a pair of $H_1$'s in the $2b2\gamma$ channel. We approach the projection using the narrow width approximation, using \texttt{HiggsTools} to estimate the gluon-fusion production cross section of $H_2$ at the HL-LHC and using CalcHEP to calculate the branching ratio $BR(H_2\rightarrow H_1H_1)$. In figure~\ref{fig:HL-LHCHeavy}, we plot the resulting estimation of $\sigma(gg\rightarrow H_2\rightarrow H_1H_1)$ as a function of the $H_2$ mass together with the sensitivity projection of HL-LHC for this process \cite{Adhikary:2018ise}. We see that most points lie below the $2\sigma$ sensitivity curve, but a few points with $m_{H_2}$ between 300 and 340~GeV might yield observable signals.  Even though the result is not shown, we also calculated this process for $H_3$ but found no relevant points. This is because the small allowed mixing between the Higgs doublet that couples to fermions and the scalar singlet yields a small top coupling of $H_3$ which suppresses the gluon-fusion production cross section of this scalar below detectable levels, even if the $H_3\rightarrow H_1 H_1$ branching ratio is significant.
We proceed to a second possibility for heavy scalars, a pseudoscalar produced in gluon-fusion production decaying into the $ZH_1$ channel and with detection in the $2b2\gamma$ channel as well. We show the obtained points in figure~\ref{fig:HL-LHCHeavy}, along with the HL-LHC sensitivity projection \cite{Adhikary:2018ise}. We find among our viable points a single instance above the sensitivity projection at $2\sigma$ with $m_A=331$~GeV.
Finally, we comment on the potential of direct observation of the charged Higgs $H^+$. In this model, as in the conventional type-I THDM, there exists the possibility of observing a light $H^+$ from top decay $t\rightarrow b H^+$, or a heavy $H^+$ produced in association with $H_1$. A novel production scenario for the heavy $H^+$ is the $Z_D$-associated production \cite{Bae:2024lov}. The presence of the $Z_D$ can also affect the decay branching ratios of the charged Higgs, potentially dominating them, i.e., $BR(H^+\rightarrow Z_DW^+)\sim1$ \cite{Davoudiasl:2014mqa,Bae:2024lov}. To our best knowledge, no dedicated study of LHC experimental sensitivity to the charged Higgs decaying in this mode has been performed. 
\begin{figure}
    \centering
    \includegraphics[width=0.48\linewidth]{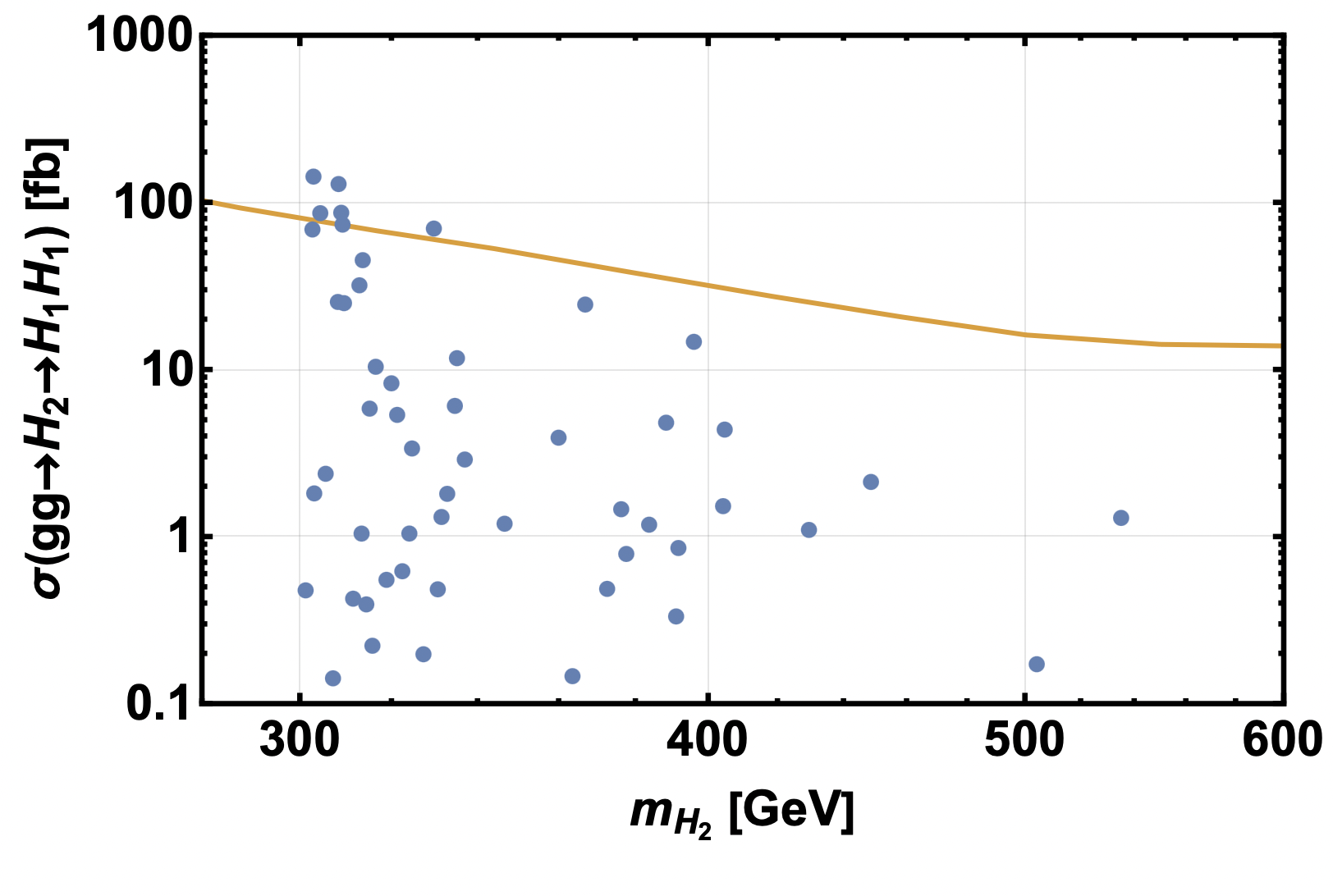}
    \includegraphics[width=0.48\linewidth]{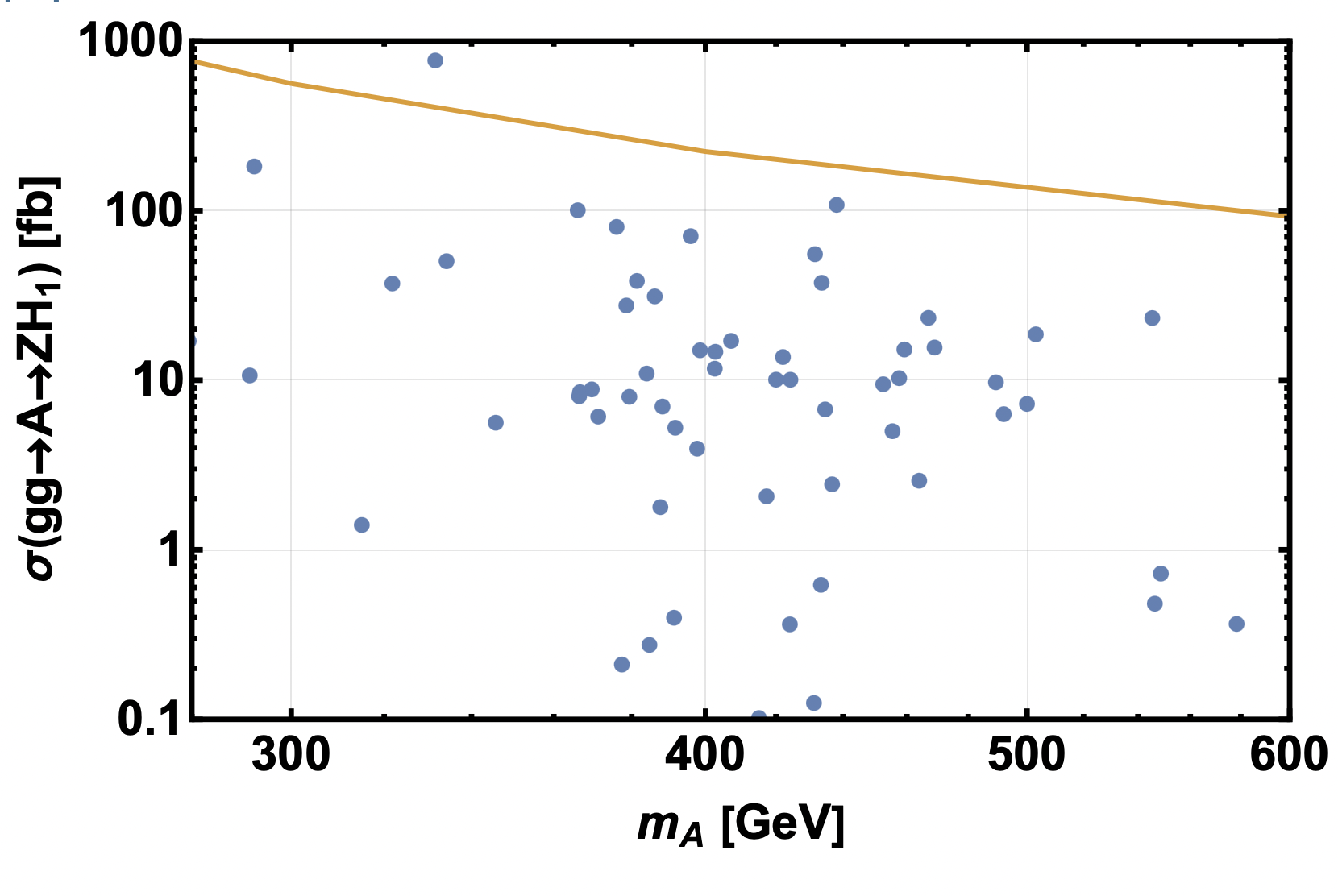}
    \caption{Projections for the observation of resonant exotic single heavy Higgs production in the $gg\rightarrow H_2\rightarrow H_1 H_1\rightarrow 2b2\gamma$ and $gg\rightarrow A\rightarrow Z H_1\rightarrow 2b2\gamma$ channels. The projections for the sensitivity of the HL-LHC at full luminosity \cite{Adhikary:2018ise} are shown in orange.} 
    \label{fig:HL-LHCHeavy}
\end{figure}
\section{Conclusions}
In this work, we have studied a scenario for WIMP DM with a dark $Z_D$ boson mixed with the SM $Z$ boson. By building the mass mixing of the neutral gauge bosons with an extra Higgs scalar doublet and a scalar singlet under the SM gauge group, we have found a viable parameter space with a rich scalar sector which can successfully drive the DM freeze-out process. In this parameter space, we have studied the collider constraints on these new scalars, together with the constraints on the interactions of the DM candidate. We have found that current constraints limit the model, but a viable region of the parameter space still exists. Within this region, we have found that there is a potential for the discovery of the new scalars, particularly for light scalars in searches for Higgs decay to two scalars in the $2b2\tau$ channel at the HL-LHC. We have also found a weaker potential for the observation of heavy scalars at the HL-LHC in the $gg\rightarrow H_2\rightarrow H_1H_1$ and $gg\rightarrow A\rightarrow Z H_1$ resonant processes in the $2b2\gamma$ channel.

\acknowledgments{This work was supported by the National Science and Technology Council under Grant No. NSTC-113-2112-M-003-007 (C.R.Chen), NSTC-111-2112-M-002-018-MY3 (C.W.Chiang), the Ministry of Education (Higher Education Sprout Project NTU-113L104022-1) (L.M.G. de la Vega), and the National Center for Theoretical Sciences of Taiwan.}
\appendix 
\section{Couplings}
\label{app:couplings}
In this appendix, we collect the couplings relevant for our calculations in this work. We introduce the notation $c_\alpha=\cos\alpha$, $s_\alpha=\sin\alpha$ for the angles $\beta$, $X\leftrightarrow\theta_X$, $W\leftrightarrow\theta_W$
\subsection{Yukawa couplings}

Here, neutral scalar couplings are normalized to the corresponding SM Higgs coupling strength, while charged Higgs couplings are given in absolute coupling strength.\\
$H_i \bar{f}f$:  \begin{equation}
\frac{(\mathcal{O}_E)_{i1}}{s_\beta}    
\end{equation} \\
$A \bar{f}f$:   \begin{equation}
    \frac{\csc\beta v_\phi}{\sqrt{v_{1}^2+v_2^2+v_\phi^2 \tan^2\beta}}
\end{equation}\\
$\text{ }H^+ \bar{u}_R d_L$:  \begin{equation}
    \frac{\sqrt{2}V_{ud}m_u\cot \beta}{v_{SM}}
\end{equation}  \\ 
$\text{ }H^+ \bar{u}_Ld_R$:  \begin{equation}
    \frac{-\sqrt{2}V_{ud}m_d\cot \beta}{v_{SM}}
\end{equation}  \\ 
$\text{ }H^+ \bar{\nu}_L e_R$: \begin{equation}
    \frac{\sqrt{2}V_{ud}m_e\cot\beta}{v_{SM}}
\end{equation}  \\ 
\subsection{Scalar-Gauge couplings}
$H_i Z Z $ :\begin{equation}
    \begin{split}
       &  \frac{1}{4} [ 2 g c_W c_X (2
   (\mathcal{O}_E)_{2i} c_\beta g_D v_{\text{SM}} s_X+\\
   & g_Y s_W c_X ((\mathcal{O}_E)_{1i} s_\beta
   v_{\text{SM}}+(\mathcal{O}_E)_{2i} c_\beta v_{\text{SM}}))+\\
   &4 g_D^2 s^2_X ((\mathcal{O}_E)_{2i} c_\beta v_{\text{SM}}+(\mathcal{O}_E)_{3i}
   v_{\phi })+\\
 &  4 (\mathcal{O}_E)_{2i} c_\beta g_D v_{\text{SM}} g_Y s_W s_X c_X+ \\
   &g^2 c^2_W c^2_X ((\mathcal{O}_E)_{1i} s_\beta
   v_{\text{SM}}+(\mathcal{O}_E)_{2i} c_\beta v_{\text{SM}})+\\
   &g_Y^2 s^2_W c^2_X ((\mathcal{O}_E)_{1i} s_\beta
   v_{\text{SM}}+(\mathcal{O}_E)_{2i} c_\beta v_{\text{SM}}) ]
    \end{split} 
\end{equation} \\
$H_i W^+ W^-$ : \begin{equation}
    \frac{1}{2} g^2 \left((\mathcal{O}_E)_{1i} s_\beta v_{\text{SM}}+(\mathcal{O}_E)_{2i} c_\beta
   v_{\text{SM}}\right)
\end{equation}\\
$H_i Z_D Z_D$ : \begin{equation}\begin{split}
  & \frac{1}{4} (4 g_D^2 c^2_X ((\mathcal{O}_E)_{2i} c_\beta
   v_{\text{SM}}+(\mathcal{O}_E)_{3i} v_{\phi })-\\
   &4 (\mathcal{O}_E)_{2i} c_\beta g_D
   v_{\text{SM}} s_X c_X (g c_W+\\
   &g_Y s_W)+s^2_X (g c_W+g_Y s_W){}^2 ((\mathcal{O}_E)_{1i} s_\beta v_{\text{SM}}+(\mathcal{O}_E)_{2i} c_\beta v_{\text{SM}}))
   \end{split}
\end{equation}\\
$H_i Z Z_D$ : \begin{equation}
    \begin{split}
        &\frac{1}{2} (s_X c_X (g_Y^2
   s^2_W ((\mathcal{O}_E)_{1i} s_\beta v_{\text{SM}}+(\mathcal{O}_E)_{2i}
   c_\beta v_{\text{SM}})-\\
   &4 g_D^2 ((\mathcal{O}_E)_{2i} c_\beta
   v_{\text{SM}}+(\mathcal{O}_E)_{3i} v_{\phi }))+g c_W
   (2 (\mathcal{O}_E)_{2i} c_\beta g_D v_{\text{SM}} s^2_X-\\
   &2 (\mathcal{O}_E)_{2i} c_\beta g_D v_{\text{SM}} c^2_X+2 g_Y s_W) s_X c_X
   ((\mathcal{O}_E)_{1i} s_\beta v_{\text{SM}}+(\mathcal{O}_E)_{2i} c_\beta
   v_{\text{SM}})+\\
   &2 (\mathcal{O}_E)_{2i} c_\beta g_D v_{\text{SM}} g_Y s_W s^2_X-2 (\mathcal{O}_E)_{2i} c_\beta g_D
   v_{\text{SM}} g_Y s_W c^2_X+\\
   &g^2 c_W s_X c_X
   ((\mathcal{O}_E)_{1i} s_\beta v_{\text{SM}}+(\mathcal{O}_E)_{2i} c_\beta
   v_{\text{SM}}))
    \end{split}
\end{equation}\\
$H_i \partial_\mu A Z^\mu$ : \begin{equation}
   \begin{split}
&\frac{-s_Xg_D\left( (\mathcal{O}_E)_{3i}s_{2\beta}v_\text{SM}+2 (\mathcal{O}_E)_{2i}s_\beta v_\phi  \right)+c_{X}v_\phi\left((\mathcal{O}_E)_{1i}c_\beta - (\mathcal{O}_E)_{2i}s_\beta   \right)(gc_W+g_Ys_W) }{\sqrt{s^2_{2\beta} v^2_{\text{SM}}+4v^2_\phi}}\\
\end{split}
\end{equation}\\
$H_i \partial_\mu A Z_D^\mu$ : \begin{equation}
   \begin{split}
&\frac{c_Xg_D\left( (\mathcal{O}_E)_{3i}s_{2\beta}v_\text{SM}+2 (\mathcal{O}_E)_{2i}s_\beta v_\phi  \right)+s_{X}v_\phi\left((\mathcal{O}_E)_{1i}c_\beta - (\mathcal{O}_E)_{2i}s_\beta   \right)(gc_W+g_Ys_W) }{\sqrt{s^2_{2\beta} v^2_{\text{SM}}+4v^2_\phi}}\\
\end{split}
\end{equation}\\
$H^+ W^- Z$ : \begin{equation}
   \begin{split}
-\frac{1}{2} g c_\beta s_\beta g_D v_{\text{SM}} s_X
\end{split}
\end{equation}\\
$H^+ W^- Z_D$ : \begin{equation}
   \begin{split}
\frac{1}{2} g c_\beta s_\beta g_D v_{\text{SM}} c_X
\end{split}
\end{equation}\\
$H^+ \partial_\mu H^- Z^\mu$ : \begin{equation}
   \begin{split}
\frac{1}{2} (g c_W c_X-2g_D c^2_\beta s_X-g_Yc_Xs_W)
\end{split}
\end{equation}\\
$H^+ \partial_\mu H^- Z_D^\mu$ : \begin{equation}
   \begin{split}
\frac{1}{2} (g c_W s_X+2g_D c^2_\beta c_X-g_Ys_Xs_W)
\end{split}
\end{equation}\\
\subsection{Scalar couplings}
    The scalar couplings which are relevant for our discussions, particularly for collider searches $H_1\rightarrow H_3H_3,AA$ and $H_2\rightarrow H_1H_1$ are\\
$H_1H_3H_3$ : \begin{equation}
   \begin{split}
1/2 &((\mathcal{O}_E)_{11} (\sqrt{2} \kappa (\mathcal{O}_E)_{23} (\mathcal{O}_E)_{33} + (\lambda_4 + \lambda_7) (\mathcal{O}_E)_{23}^2 v_{1} +  \lambda_6 (\mathcal{O}_E)_{33}^2 v_{1})\\
&+ 6 \lambda_2 (\mathcal{O}_E)_{21} (\mathcal{O}_E)_{23}^2 v_{2} + 
   \lambda_5 (\mathcal{O}_E)_{33} (2 (\mathcal{O}_E)_{23} (\mathcal{O}_E)_{31} + (\mathcal{O}_E)_{21} (\mathcal{O}_E)_{33}) v_{2} +\\
   &6 \lambda_3 (\mathcal{O}_E)_{31} (\mathcal{O}_E)_{33}^2 v_\phi + 
   \lambda_5 (\mathcal{O}_E)_{23} ((\mathcal{O}_E)_{23} (\mathcal{O}_E)_{31} + 2 (\mathcal{O}_E)_{21} (\mathcal{O}_E)_{33}) v_\phi + \\
  & (\mathcal{O}_E)_{13}^2 (6 \lambda_1 (\mathcal{O}_E)_{11} v_{1} + (\lambda_4 + \lambda_7) (\mathcal{O}_E)_{21} v_{2} + 
      \lambda_6 (\mathcal{O}_E)_{31} v_\phi) + \\
  & (\mathcal{O}_E)_{13} (\sqrt{2} \kappa (\mathcal{O}_E)_{23} (\mathcal{O}_E)_{31} + \sqrt{2} \kappa (\mathcal{O}_E)_{21} (\mathcal{O}_E)_{33} + \\
      &2 (\lambda_4 + \lambda_7) (\mathcal{O}_E)_{23} ((\mathcal{O}_E)_{21} v_{1} + (\mathcal{O}_E)_{11} v_{2}) + 
      2 \lambda_6 (\mathcal{O}_E)_{33} ((\mathcal{O}_E)_{31} v_{1} + (\mathcal{O}_E)_{11} v_\phi)))
\end{split}
\end{equation}\\
  $H_2H_1H_1$ : \begin{equation}
   \begin{split}
   1/2 &((\mathcal{O}_E)_{11} (\sqrt{2} \kappa (\mathcal{O}_E)_{22} (\mathcal{O}_E)_{31} + 2 (\lambda_4 + \lambda_7) (\mathcal{O}_E)_{21} (\mathcal{O}_E)_{22} v_{1} +\\ 
      &(\mathcal{O}_E)_{32} (\sqrt{2} \kappa (\mathcal{O}_E)_{21} + 2 \lambda_6 (\mathcal{O}_E)_{31} v_{1})) + 
   6 \lambda_2 (\mathcal{O}_E)_{21}^2 (\mathcal{O}_E)_{22} v_{2} +\\
   &\lambda_5 (\mathcal{O}_E)_{31} ((\mathcal{O}_E)_{22} (\mathcal{O}_E)_{31} + 2 (\mathcal{O}_E)_{21} (\mathcal{O}_E)_{32}) v_{2} + 
   6 \lambda_3 (\mathcal{O}_E)_{31}^2 (\mathcal{O}_E)_{32} v_\phi +\\
   &\lambda_5 (\mathcal{O}_E)_{21} (2 (\mathcal{O}_E)_{22} (\mathcal{O}_E)_{31} + (\mathcal{O}_E)_{21} (\mathcal{O}_E)_{32}) v_\phi +
    (\mathcal{O}_E)_{11}^2 ((\lambda_4 + \lambda_7) (\mathcal{O}_E)_{22}\\
    &v_{2} + \lambda_6 (\mathcal{O}_E)_{32} v_\phi) + 
   (\mathcal{O}_E)_{12} (\sqrt{2} \kappa (\mathcal{O}_E)_{21} (\mathcal{O}_E)_{31} + 
      6 \lambda_1 (\mathcal{O}_E)_{11}^2 v_{1} +\\
      &(\lambda_4 + \lambda_7) (\mathcal{O}_E)_{21}^2 v_{1} + 
      2 (\lambda_4 + \lambda_7) (\mathcal{O}_E)_{11} (\mathcal{O}_E)_{21} v_{2} + 
      \lambda_6 (\mathcal{O}_E)_{31} ((\mathcal{O}_E)_{31} v_{1} +\\
      &2 (\mathcal{O}_E)_{11} v_\phi)))
   \end{split}
\end{equation}\\
 $H_1AA$ : \begin{equation}
   \begin{split}
   &(4 v_\phi^2 ((\lambda_4 + \lambda_7) (\mathcal{O}_E)_{11} v_{1} + 2 \lambda_2 (\mathcal{O}_E)_{21} v_{2} + 
      \lambda_5 (\mathcal{O}_E)_{31} v_\phi) s_\beta^2 + \\
&   8 \lambda_1 (\mathcal{O}_E)_{11} v_{1} v_\phi^2 c_\beta^2 + 
   4 \lambda_6 (\mathcal{O}_E)_{31} v_\phi^3 c_\beta^2 + 
   \lambda_6 (\mathcal{O}_E)_{11} v_{1} v_{\text{SM}}^2 s_{2\beta}^2 + \\
   &2 \lambda_3 (\mathcal{O}_E)_{31} v_\phi v_{\text{SM}}^2 s_{2\beta}^2 + 
   2 \sqrt{2}
     \kappa v_\phi s_\beta (-2 (\mathcal{O}_E)_{31} v_\phi c_\beta + (\mathcal{O}_E)_{11} v_{\text{SM}} s_{2\beta}) + \\
&   (\mathcal{O}_E)_{21} (4 (\lambda_4 + \lambda_7) v_{2} v_\phi^2 c_\beta^2 + 
      2 \sqrt{2} \kappa v_\phi v_{\text{SM}} c_\beta s_{2\beta} + 
      \lambda_5 v_{2} v_{\text{SM}}^2 s_{2\beta}^2))/(8 v_\phi^2 + 
   2 v_{\text{SM}}^2 s_{2\beta}^2)
   \end{split}
\end{equation}\\
\bibliography{main}
\bibliographystyle{unsrtnat}
\end{document}